\documentclass[12pt, a4paper] {article}

\usepackage{amsfonts}
\usepackage{verbatim}
\usepackage{color}
\usepackage{amsmath}
\usepackage{graphicx}
\usepackage{bm}
\usepackage{amssymb}
\usepackage{latexsym}
\usepackage{mathrsfs}
\usepackage{textcomp}
\usepackage{ulem}
\begin{document}

\begin{center}
{\bf\large{Particle on a torus knot: Symplectic analysis} }

\vskip 1.5 cm

{\sf{ \bf Anjali S and Saurabh Gupta}}\\
\vskip .1cm
{\it Department of Physics, National Institute of Technology Calicut,\\ Kozhikode - 673 601, Kerala, India}\\
\vskip .15cm
{E-mails: {\tt anjalisujatha28@gmail.com, saurabh@nitc.ac.in}}
\end{center}
\vskip 1cm

\noindent
{\bf Abstract:}
We quantize a particle confined to move on a torus knot satisfying constraint condition ($p\theta+q\phi) \approx 0$, within the context of a geometrically motivated approach – the Faddeev-Jackiw formalism. We also deduce the constraint spectrum and discern the basic brackets of the theory. We further reformulate the original gauge non-invariant theory into a {\it{physically}} equivalent gauge theory, which is free from any additional Wess-Zumino variables, by employing symplectic gauge invariant formalism. In addition, we analyze the reformulated gauge invariant theory within the framework of BRST formalism to establish the off-shell nilpotent and absolutely anti-commuting (anti-)BRST symmetries. Finally, we construct the conserved (anti-)BRST charges which satisfy the physicality criteria and turn out to be the generators of corresponding symmetries.

\vskip 1.5 cm



\newpage
\section{Introduction}
\label{intro}
Knot theory plays a key role in a wide range of physical scenarios ranging from string theory, statistical physics, polymer physics to various studies in molecular biology \cite{wit,sum,fadd}. It establishes a connection to the exploration of algebraic structure inherent in renormalizable quantum field theories \cite{knt1}. Knots are also expected to determine gauge invariant observables in some quantum gravity approaches \cite{knt_rovelli}. Consequently, some generalizations of knot theory have been employed to construct operator algebra which may express quantum geometry at non-perturbative level \cite{knt_smolin}. The classical and quantum mechanics of a free particle constrained to move on a knot that winds a torus have been explored to deduce the energy spectrum \cite{int2} and Hamiltonian analysis has been carried out to account for dynamics and symmetries \cite{1}. The Becchi-Rouet-Stora-Tyutin (BRST) and anti-BRST symmetries have been established via Batalin-Fradkin-Vilkovisky formalism \cite{int3}. Moreover, the  problem of a charge moving in a torus knot in the presence of a uniform magnetic field has been explored in a semi-classical quantization procedure \cite{2}.

As long as the conventional canonical quantization process is concerned, it is not applicable in a straightforward manner to the systems with singular Lagrangian or constrained systems \cite{int5,gauge_Henn}. To deal with it, Anderson-Bergmann and Dirac independently developed an approach for a consistent and systematic treatment of constraints \cite{int7,int9,int6}. Dirac's formalism introduces an effective procedure for deducing all constraints in a theory and at the same time classifies them into primary, secondary, tertiary {\it{etc.}} and further into first-class and second-class. For second-class constrained systems, Dirac introduced a generalized bracket known as Dirac bracket \cite{int6}. However, the procedure of finding Dirac brackets is found to be convoluted. Alternatively, Faddeev and Jackiw developed a geometrically motivated methodology in dealing with constrained systems which makes use of the symplectic nature of phase space \cite{3}. In Faddeev-Jackiw formalism, the singular nature of symplectic two-form matrix predicts the existence of constraints and they are procured in an iterative fashion ({\it{cf.}} for details \cite{4,int10,FJ1,FJ3}). This formalism is further modified to incorporate Dirac-Bergmann algorithm \cite{mFJ2,mFJ3} and implemented in several systems such as anti self-dual Yang-Mills equations \cite{modFJ3}, Abelian and non-Abelian exotic action for garvity \cite{modFJ_grav}, four-dimensional BF theory \cite{modFJ_BF},  and Christ-Lee model \cite{modFJ4}.

It is a well-known fact that the quantization of a gauge non-invariant theory (system endowed with second-class constraints) confronts the problem of factor ordering \cite{kuch,sec_oprtr}. Consequently, numerous approaches have been developed to construct an equivalent gauge invariant version of a gauge non-invariant theory.  
To mention some approaches, Batalin-Fradkin-Fradkina-Tyutin proposed a reformulation technique by enlarging the phase space with the introduction of Wess-Zumino variables \cite{bfft2,bfft3}. Further, for the system of solely second-class constraints, Mitra and Rajaraman developed a methodology to construct gauge theory with a subset of half of the constraints having mutual vanishing Poisson brackets among them \cite{mit1,mit2}. In addition, another formalism is introduced which deforms the structure of local symmetries of Lagrangian formulation in the conversion procedure \cite{conv.gau}. Furthermore, a methodology based on symplectic formalism is also proposed where an arbitrary function depending upon the phase space variables and Wess-Zumino variable is introduced to reformulate a gauge non-invariant theory \cite{5,symplec,sym1}.

On the other hand, gauge theories - endowed with the first-class constraints in the language of Dirac's prescription for the classification of constraints - play a crucial role in theoretical physics as they capacitate the investigations of fundamental interactions. Quantization of such theories requires special attention due to the presence of superfluous degrees of freedom. 
The BRST formalism is one of the most favorite schemes of quantization for a gauge theory \cite{brst1,brst3,malik}.  In BRST formalism, the Hilbert space of gauge theory is enlarged and the notion of gauge invariance is replaced by BRST invariance which contains anti-commuting variables - Faddeev-Popov ghosts (and corresponding anti-ghosts). The resulting theory possesses a more generalized gauge invariance, known as BRST invariance, where the effect of gauge fixing is achieved without breaking the BRST symmetry \cite{rotor}.

Keeping the above in mind, our first motive of present investigation is to explore the constraint structure and subsequently quantize the system of particle on a torus knot via a geometrically motivated approach - the Faddeev-Jackiw formalism. The second motive is to reformulate the system as a gauge theory by employing symplectic gauge invariant formalism and disclose hidden gauge symmetries. Additionally, as a final motive, we wish to deduce the off-shell nilpotent and absolutely anti-commuting (anti-)BRST symmetries for this system.

The contents of the paper are organized as follows. Section 2 gives a brief preface about a particle restricted to move on a torus knot. We discuss the constraint structure and Faddeev-Jackiw quantization in section 3. Our section 4 comprises the reformulation of the system as a gauge theory via symplectic gauge invariant formalism and gives an account of the gauge symmetries. Our section 5 contains the derivation of off-shell nilpotent and absolutely anti-commuting (anti-)BRST transformations and corresponding conserved charges. Finally, we summarize our results in section 6. 

Our Appendix A consists of explicit computation of Dirac brackets and Appendix B deals with the calculation of equations of motion. 

\section{Preliminaries}
\label{sec:1}
We begin with toroidal coordinates $(\eta, \theta, \phi)$ which are related with Cartesian coordinates $(x, y, z)$ in the following fashion \cite{knot2}
\begin{equation}
	x = \frac{a\sinh\eta\cos\phi}{\cosh\eta - \cos\theta}, \quad y = \frac{a\sinh\eta\sin\phi}{\cosh\eta - \cos\theta}, \quad z = \frac{a\sin\theta}{\cosh\eta - \cos\theta},
\end{equation}
where the coordinates ranges from $0\leq \eta < \infty$, $-\pi < \theta \leq +\pi$ and $0 \leq \phi < 2\pi$. The parameter $a$ is defined as $a^{2} = R^{2} - r^{2}$, where $R$ is major radius and $r$ is minor radius of the torus. Moreover, the toroidal coordinates satisfy following relations
\begin{equation}
 \tan\phi = \displaystyle\frac{y}{x}, \quad x^{2}+y^{2} + \left(z-a\cot\theta\right)^{2} = \frac{a^{2}}{\sin^{2}\theta}, \quad x^{2}+y^{2}+z^{2}+a^{2} = \displaystyle2a\left(x^{2}+y^{2}\right)^{\frac{1}{2}}\coth\eta,
\end{equation}
where, $\eta$ taking a constant value represents a toroidal surface,  $\theta$ being a constant generates spherical bowls and a constant value of $\phi$ gives half planes \cite{knot2,knot3}.  Here we are interested in toroidal surface and coordinate $\eta$ taking a specific value, say {\it{e.g.}} $\eta_{c} = \cosh^{-1}(\frac{R}{r})$ gives a toroidal surface \cite{int2}. A larger value of $\eta_{c}$ generates a torus of smaller thickness.

In our present investigation, we consider the case of a particle constrained to move on a torus knot.  A torus  knot is defined to be a knot that resides on an unknotted torus without crossing under or over themselves on the torus surface. Every torus knot is a $(p, q)$-torus knot for some pair of integers, where $p$ and $q$ are relatively prime. Here the torus knot wraps $p$ times along the meridian curve and $q$ times along the longitudinal curve ({\it{cf.}} \cite{knot} for details). 

In order to obtain the case of a particle restricted to move on a knot that winds the torus, we impose the constraint condition $p\theta+q\phi \approx 0$ into the system, where $p$ and $q$ are mutually prime numbers \cite{1,2}. Thus, the Lagrangian describing a particle of mass $m$ constrained to move on a $(p, q)$-torus knot is given as
\begin{equation} \label{L}
L \;=\; \frac{ma^2}{2(\cosh\eta - \cos\theta)^2}\left[  \dot{\eta^2} + \dot{\theta^2} + \sinh^2\eta \; \dot{\phi^2} \right] + \lambda(p\theta+q\phi),
\end{equation}
where $\lambda$ represents the Lagrange multiplier and overdot indicates time derivative. The canonical momenta corresponding to variables $\eta$, $\theta$, $\phi $ and $\lambda$ are, respectively, listed below
\begin{eqnarray}
P_\eta = \displaystyle\frac{ma^2\dot{\eta}}{(\cosh\eta - \cos\theta)^2}, \quad 
P_\theta = \frac{ma^2\dot{\theta}}{(\cosh\eta - \cos\theta)^2}, \quad P_\phi = \displaystyle\frac{ma^2 \sinh^2\eta \dot{\phi}}{(\cosh\eta - \cos\theta)^2 }, \quad P_{\lambda} = 0.
\end{eqnarray} 
The canonical Hamiltonian corresponding to the Lagrangian \eqref{L} takes the following form
\begin{equation} \label{H}
H \;=\; \frac{(\cosh\eta - \cos\theta)^2}{2ma^2}\left[ P_\eta^2 + P_\theta^2 + \frac{P_{\phi}^2}{\sinh^2 \eta} \right] - \lambda(p\theta+q\phi).
\end{equation}
In the context of Dirac formalism, we infer the primary constraint in the system as 
\begin{equation}
	\varphi_{1} \; \equiv \; P_{\lambda} \; \approx \; 0.
\end{equation}
The consistency condition of $\varphi_{1}$ leads to the secondary constraint $\varphi_{2}$ as
\begin{equation}
	\dot{\varphi}_{1} \; = \; \left\lbrace \varphi_{1}, H_{p} \right\rbrace \approx 0 \quad \Longrightarrow \quad \varphi_{2} \; \equiv \; (p\theta+q\phi) \approx 0,
\end{equation}
where the explicit form of primary Hamiltonian $H_{p}$ is given by
\begin{equation} \label{primary_H}
	H_{p} \; = \; \frac{(\cosh\eta - \cos\theta)^2}{2ma^2}\left[ P_\eta^2 + P_\theta^2 + \frac{P_{\phi}^2}{\sinh^2 \eta} \right] - \lambda(p\theta+q\phi) + \upsilon P_{\lambda}, 
\end{equation}
with $\upsilon$ is a Lagrange multiplier. The consistency condition of secondary constraint $\varphi_{2}$ leads to the tertiary constraint $\varphi_{3}$, as 
\begin{equation}
	\dot{\varphi}_{2} \; = \; \left\lbrace \varphi_{2}, H_{p} \right\rbrace \approx 0  \Longrightarrow  \varphi_{3} \; \equiv \;  \frac{(\cosh\eta - \cos\theta)^{2}}{ma^{2}} \left( pP_\theta + \frac{qP_{\phi}}{\sinh^{2}\eta} \right) \approx 0.
\end{equation}
There are no further constraints present. 
Among these three constraints, $\varphi_{1}$ has vanishing Poisson brackets with $\varphi_{2}$ and $\varphi_{3}$ whereas the Poisson bracket between $\varphi_{2}$ and $\varphi_{3}$ is given as $\lbrace \varphi_{2}, \varphi_{3}\rbrace = \frac{(\cosh\eta - \cos\theta)^{2}}{ma^{2}} \left( p^{2} + \frac{q^{2}}{\sinh^{2}\eta} \right) $. Thus, $\varphi_{1}$ is first-class, at the same time, $\varphi_{2}$ and $\varphi_{3}$ are second-class constraints.

\section{Faddeev-Jackiw formalism}
\label{sect:2}
In this section, we investigate the system of particle on a torus knot in the context of Faddeev-Jackiw formalism \cite{3,4}. We also realize the constraint structure and furnish all the basic brackets existing in the system.
\subsection{ A concise recipe}
\label{sect:2.1}
To begin with, we consider the first-order Lagrangian where no quadratic terms of generalized velocities are present. 
The general form of the first-order Lagrangian can be  given as 
\begin{equation}
	L^{(0)} = a_{i}^{(0)}\dot{\zeta}^{(0)}_{i} - V^{(0)}(\zeta),
\end{equation}
where $ a_{i}^{(0)}$ and $V^{(0)}(\zeta)$ represent the canonical one-form and symplectic potential, respectively and the subscript $i$ denotes the dimension of extended configuration space. In this formalism, the geometric structures are obtained with the aid of symplectic two-form $f^{(0)}_{ij}$, defined as the exterior product of canonical one-form \cite{4}. The equations of motion are derived in terms of symplectic matrix $f_{ij}^{(0)}$ in the following manner 
\begin{equation} \label{f_ij_eqm}
f_{ij}^{(0)}\dot{\zeta}^{(0)j} \;=\; \frac{\partial V^{(0)}(\zeta)}{\partial \zeta^{(0)i}},
\end{equation}
where the symplectic matrix is defined by
\begin{eqnarray} \label{f_ij}
f_{ij}^{(0)} &=& \frac{\partial a^{(0)}_{j}(\zeta)}{\partial \zeta^{(0)i}}- \frac{\partial a^{(0)}_{i}(\zeta)}{\partial \zeta^{(0)j}}. 
\end{eqnarray}
The (non-)singular nature of this symplectic matrix determines whether the system under investigation is  (un)constrained. In Faddeev-Jackiw formalism, the constraints are procured in an iterative fashion from the zero-modes of singular symplectic matrix. Successive incorporation of all constraints into the canonical sector of first-order Lagrangian yields a non-singular matrix (in case of gauge non-invariant theory). Further, from the inverse of non-singular symplectic matrix, the basic brackets are deduced  ({\it{cf.}} \cite{4,int10} for details). 

\subsection{A particle on a torus knot}
\label{sect:2.2}

Considering the case of a particle on a torus knot and taking the canonical momenta as the auxiliary variables, we construct the first-order Lagrangian as 
\begin{equation} \label{L_0}
L_f^{(0)} \;=\;  \dot{\eta} P_\eta + \dot{\theta}  P_\theta+ \dot{\phi}  P_\phi- V^{(0)},
\end{equation}
where the symplectic potential $V^{(0)}$ is given by
\begin{equation}
V^{(0)} \;=\; \frac{(\cosh\eta - \cos\theta)^2}{2ma^2}\left[ P_\eta^2 + P_\theta^2 + \frac{P_{\phi}^2}{\sinh^2 \eta} \right] - \lambda(p\theta+q\phi).
\end{equation}
 In order to deduce the constraint structure  we calculate the form of zeroth-iterated symplectic matrix $f_{ij}^{(0)}$. To start with, we choose the set of symplectic variables as $\zeta^{(0)} \;=\; \left\{\eta, P_\eta, \theta, P_\theta, \phi, P_\phi, \lambda\right\}.$
The corresponding components of canonical one-forms are listed below:
\begin{eqnarray}
&a^{(0)}_{\eta}\;=\;P_\eta,  \quad a^{(0)}_{\theta}\;=\;P_{\theta}, \quad a^{(0)}_{\phi}\;=\;P_{\phi}, \quad
 a^{(0)}_{P_{\eta}} \;=\; a^{(0)}_{P_{\theta}} \;=\; a^{(0)}_{P_{\phi}} \;=\; a^{(0)}_{\lambda} \;=\; 0.&
\end{eqnarray}
We obtain the following form of symplectic matrix $f_{ij}^{(0)}$ as
\begin{eqnarray}
f_{ij}^{(0)} \;=\;
\begin{pmatrix}
0 & -1 &0 &0 &0 &0 &0\\
1 &0 &0 &0 &0 &0 &0\\
0 &0 &0 &-1 &0 &0 &0\\
0 &0 &1 &0 &0 &0 &0\\
0 &0 &0 &0 &0 &-1 &0\\
0 &0 &0 &0 &1 &0 &0\\
0 &0 &0 &0 &0 &0 &0
\end{pmatrix},
\end{eqnarray}
which is a singular matrix. This indicates the system under study is a constrained system. The zero-mode of $f_{ij}^{(0)}$ is $(\nu^{(0)})^{T} \;=\; 
	(0 \; 0 \; 0 \; 0 \; 0 \; 0 \; \nu_\lambda)$, where $\nu_\lambda$ is an arbitrary constant. To obtain the constraints, we contract this zero-mode with Euler-Lagrange equations ({\it{cf.}} \eqref{f_ij_eqm}) as illustrated below
\begin{eqnarray} \label{C_0}
\Omega^{(0)} \;=\; (\nu^{(0)})^{T}\frac{\partial V^{(0)}(\zeta)}{\partial \zeta^{(0)}} \;=\; 0 \quad \Longrightarrow \quad \Omega^{(0)} \;=\; \nu_{\lambda}(p\theta+q\phi) \;=\; 0. 
\end{eqnarray} 
Now we incorporate the constraint $\Omega^{(0)}$ into the canonical sector of first-order Lagrangian \eqref{L_0} with the help of Lagrange multiplier. The form of first-iterated Lagrangian is given as
\begin{equation} \label{L_1}
L_{f}^{(1)} \;=\; \dot{\eta} P_\eta + \dot{\theta}  P_\theta+ \dot{\phi}  P_\phi + \dot{\beta} \left(p\theta+q\phi\right)  - V^{(1)},
\end{equation}
where  $\beta$ denotes a Lagrange multiplier. Here the form of symplectic potential $V^{(1)}$ is given by
\begin{equation}
V^{(1)} \;=\; V^{(0)}\vert_{\Omega^{(0)}=0} \;=\; \frac{(\cosh\eta - \cos\theta)^2}{2ma^2}\left[ P_\eta^2 + P_\theta^2 + \frac{P_{\phi}^2}{\sinh^2 \eta} \right].
\end{equation}
The set of first-iterated symplectic variables becomes $\zeta^{(1)} \;=\; \{\eta, P_\eta, \theta, P_\theta, \phi, P_\phi, 
\beta\}$ and the corresponding canonical one-forms are given as
\begin{eqnarray}
a^{(1)}_{\eta} \;=\; P_\eta, \quad  a^{(1)}_{\theta}=P_{\theta}, \quad a^{(1)}_{\phi}=P_{\phi}, \quad a^{(1)}_{\beta} = p\theta+q\phi, \quad 
a^{(1)}_{P_{\eta}} \;=\; a^{(1)}_{P_{\theta}} \;=\; a^{(1)}_{P_{\phi}} \;=\; 0.
\end{eqnarray}
With this the first-iterated symplectic matrix $f_{ij}^{(1)}$ turns out to be 
\begin{eqnarray}
f_{ij}^{(1)} \;=\;
\begin{pmatrix}
0 & -1 &0 &0 &0 &0 &0\\
1 &0 &0 &0 &0 &0 &0\\
0 &0 &0 &-1 &0 &0 &p\\
0 &0 &1 &0 &0 &0 &0\\
0 &0 &0 &0 &0 &-1 &q\\
0 &0 &0 &0 &1 &0 &0\\
0 &0 &-p &0 &-q &0 &0
\end{pmatrix}.
\end{eqnarray}
We observe the symplectic matrix $f_{ij}^{(1)}$ is singular and possesses zero-mode of order $1 \times 7$. This zero-mode is given as $(\nu^{(1)})^{T} \;=\; (0 \;0 \;0 \;p \;0 \;q \;1)$. The contraction of this zero-mode with the gradient of first-iterated symplectic potential, {\it{i.e.}}
\begin{equation} 
\Omega^{(1)} \;=\; (\nu^{(1)})^{T}\frac{\partial V^{(1)}(\zeta)}{\partial \zeta^{(1)}} \;=\; 0,
\end{equation}
 in turn, generates the next constraint ($\Omega^{(1)}$) as
\begin{equation} \label{C_1}
\Omega^{(1)} \;=\; \frac{(\cosh\eta - \cos\theta)^{2}}{ma^{2}} \left( pP_\theta + \frac{qP_{\phi}}{\sinh^{2}\eta} \right) \;=\; 0.
\end{equation}
We now incorporate the constraint $\Omega^{(1)}$  into the first-iterated Lagrangian \eqref{L_1} with the help of a Lagrange multiplier, say $\rho$, to construct the second-iterated Lagrangian as
\begin{eqnarray} \label{L_2}
L_{f}^{(2)} \;=\; \dot{\eta} P_\eta + \dot{\theta}  P_\theta+ \dot{\phi}  P_\phi + \dot{\beta} \left(p\theta+q\phi\right) 
+ \dot{\rho} \frac{(\cosh\eta - \cos\theta)^{2}}{ma^{2}} \left( pP_\theta + \frac{qP_{\phi}}{\sinh^{2}\eta} \right)  - V^{(2)},
\end{eqnarray}
where the form of symplectic potential $V^{(2)}$ is 
\begin{equation}
V^{(2)} \;=\; V^{(1)}\vert_{\Omega^{(1)}=0} = \frac{(\cosh\eta - \cos\theta)^2}{2ma^2}\left[ P_\eta^2 + P_\theta^2 + \frac{P_{\phi}^2}{\sinh^2 \eta} \right].
\end{equation}
To construct the second-iterated symplectic matrix, we choose the set of second-iterated symplectic variables as $\zeta^{(2)} \;=\; \left\{\eta, P_\eta, \theta, P_\theta, \phi, P_\phi, \beta, \rho \right\}$ 
and identifying corresponding canonical one-forms as
\begin{eqnarray}
&a^{(2)}_{\eta}\;=\; P_\eta, \quad  a^{(2)}_{\theta} \;=\; P_{\theta}, \quad a^{(2)}_{\phi} \;=\; P_{\phi},  \quad a^{(2)}_{\beta} \;=\; p\theta+q\phi,& \\ \nonumber
& \displaystyle a^{(2)}_{\rho} \;=\; \frac{(\cosh\eta - \cos\theta)^{2}}{ma^{2}} \left( pP_\theta + \frac{qP_{\phi}}{\sinh^{2}\eta} \right), \quad a^{(2)}_{P_{\eta}} \;=\; a^{(2)}_{P_{\theta}} \;=\; a^{(2)}_{P_{\phi}} \;=\; 0.&
\end{eqnarray}
Thus, the second-iterated symplectic matrix $f_{ij}^{(2)}$ looks like
\begin{eqnarray}
f_{ij}^{(2)} \;=\;
\begin{pmatrix}
0 & -1 &0 &0 &0 &0 &0 &f_{\eta\rho}\\
1 &0 &0 &0 &0 &0 &0 &0\\
0 &0 &0 &-1 &0 &0 &p &f_{\theta\rho}\\
0 &0 &1 &0 &0 &0 &0 &f_{P_{\theta}\rho}\\
0 &0 &0 &0 &0 &-1 &q &0\\
0 &0 &0 &0 &1 &0 &0 &f_{P_{\phi}\rho}\\
0 &0 &-p &0 &-q &0 &0 &0\\
-f_{\eta\rho} &0 &-f_{\theta\rho} &-f_{P_{\theta}\rho} &0 &-f_{P_{\phi}\rho} &0 &0
\end{pmatrix},
\end{eqnarray}
where  the explicit form of entries $f_{\theta\rho}$, $f_{P_{\phi}\rho}$, $f_{P_{\theta}\rho}$ and $f_{\eta\rho}$ are given as
\begin{eqnarray} \label{f_{ij}_forms}
&& \displaystyle f_{\theta\rho} = \frac{2(\cosh\eta - \cos\theta)\sin\theta}{ma^{2}} \left(pP_{\theta} + \frac{qP_{\phi}}{\sinh^{2}\eta} \right), 
  \\ \nonumber
&&\displaystyle f_{P_{\phi}\rho} = \frac{(\cosh\eta-\cos\theta)^{2}}{ma^{2}} \frac{q}{\sinh^{2}\eta}, \quad \displaystyle f_{P_{\theta}\rho} = \frac{(\cosh\eta-\cos\theta)^{2}p}{ma^{2}}, \\ \nonumber
&&  \displaystyle f_{\eta\rho} = \frac{2(\cosh\eta - \cos\theta)}{ma^{2}} \left[p\sinh\eta P_{\theta} + \frac{qP_{\phi}}{\sinh\eta} - \frac{(\cosh\eta - \cos\theta)qP_{\phi}\cosh\eta}{\sinh^{3}\eta} \right].
\end{eqnarray}
It is worth mentioning that the second-iterated symplectic matrix $f_{ij}^{(2)}$ is found to be non-singular. So,
it does not have any zero-mode and we can conclude that we have obtained all the constraints in the system. This, in turn, indicates that we can stop the iterative process and procure all the basic brackets in the theory. 

At this juncture we would like to point out  a key feature of constrained systems, when explored  within the framework of Faddeev-Jackiw formalism, if the symplectic matrix is singular even after incorporating all the constraints into the canonical sector then it hints the presence of some gauge symmetry. Accordingly, a suitable gauge condition has to be chosen in order to get a non-singular symplectic matrix \cite{int10,modFJ4}.

\subsection{Symplectic quantization: Basic brackets}
\label{sect:2.3}
 As we have already mentioned that the presence of a non-singular symplectic matrix indicates that our system is gauge non-invariant. Thus, we can compute the inverse and in turn procure all basic brackets in the theory from its components. The inverse of the symplectic matrix $f_{ij}^{(2)}$ is calculated as
\begin{eqnarray}
(f_{ij}^{(2)})^{-1} =
\begin{pmatrix}
0  & 1 & 0 & 0 & 0 & 0 & 0 & 0\\
-1 & 0 & 0 &\frac{pf_{\eta\rho}}{c} & 0 &\frac{qf_{\eta\rho}}{c} & \frac{f_{\eta\rho}}{c} & 0\\
0   &0 & 0 &\frac{qf_{P_{\phi}\rho}}{c} &0 &-\frac{qf_{P_{\theta}\rho}}{c} &-\frac{f_{P_{\theta}\rho}}{c} & 0\\
0  &-\frac{pf_{\eta\rho}}{c} &-\frac{qf_{P_{\phi}\rho}}{c} &0 &\frac{pf_{P_{\phi}\rho}}{c} &\frac{qf_{\theta\rho}}{c} &\frac{f_{\theta\rho}}{c} &-\frac{p}{c}\\
0 & 0 & 0 &-\frac{pf_{P_{\phi}\rho}}{c} &0 &\frac{pf_{P_{\theta}\rho}}{c} &-\frac{f_{P_{\phi}\rho}}{c} &0\\
0 &-\frac{qf_{\eta\rho}}{c} &\frac{qf_{P_{\theta}\rho}}{c} &-\frac{qf_{\theta\rho}}{c} &-\frac{pf_{P_{\theta}\rho}}{c} &0 &0 &-\frac{q}{c}\\
0 &-\frac{f_{\eta\rho}}{c} &\frac{f_{P_{\theta}\rho}}{c} &-\frac{f_{\theta\rho}}{c} &\frac{f_{P_{\phi}\rho}}{c} &0 &0 &-\frac{1}{c}\\
0 &0 &0 &\frac{p}{c} &0 &\frac{q}{c} &\frac{1}{c} &0
\end{pmatrix},
\end{eqnarray}
where $c = \displaystyle\frac{(\cosh\eta - \cos\theta)^{2}}{ma^{2}} \left( p^{2} + \frac{q^{2}}{\sinh^{2} \eta} \right)$. So, the basic brackets can be obtained with the help of following relationship
\begin{equation}
	\lbrace \zeta_{i}^{(2)}, \zeta_{j}^{(2)} \rbrace = (f_{ij}^{(2)})^{-1}.
\end{equation}
The generalized basic brackets in the theory are given, as below
\begin{eqnarray} \label{FJ_basic brac}
	&& \lbrace\eta, P_{\eta}\rbrace = 1, \quad  \lbrace \theta, P_{\theta} \rbrace = \displaystyle \frac{q^{2}}{(p^{2}\sinh^{2}\eta+q^{2})}, \quad  \lbrace \theta, P_{\phi} \rbrace = \displaystyle -\frac{qp\sinh^{2}\eta}{(p^{2}\sinh^{2}\eta+q^{2})},  \\ \nonumber 
	&& \lbrace P_{\eta}, P_{\theta}\rbrace = \displaystyle \frac{2p}{(p^{2}\sinh^{2}\eta+q^{2})}\left[ \frac{\sinh\eta}{(\cosh\eta-\cos\theta)} \left(p\sinh^{2}\eta P_{\theta}+qP_{\phi}\right) -q\coth\eta P_{\phi} \right],\\ \nonumber 
	&&  \lbrace P_{\eta}, P_{\phi}\rbrace = \displaystyle \frac{2q}{(p^{2}\sinh^{2}\eta+q^{2})}\left[ \frac{\sinh\eta}{(\cosh\eta-\cos\theta)} \left(p\sinh^{2}\eta P_{\theta}+qP_{\phi}\right) -q\coth\eta P_{\phi} \right], \\ \nonumber
	&& \lbrace P_{\eta}, \beta \rbrace =\displaystyle \frac{2}{(p^{2}\sinh^{2}\eta+q^{2})}\left[ \frac{\sinh\eta}{(\cosh\eta-\cos\theta)} \left(p\sinh^{2}\eta P_{\theta}+qP_{\phi}\right) -q\coth\eta P_{\phi} \right],  \\ \nonumber
	\end{eqnarray} 
	\begin{eqnarray*}
	&&\lbrace P_{\theta}, P_{\phi}\rbrace =\displaystyle \frac{2q \sin\theta}{( p^{2}\sinh^{2}\eta+q^{2})}\frac{\left(p\sinh^{2}\eta P_{\theta}+qP_{\phi}\right)}{(\cosh\eta-\cos\theta)}, 
	\quad \lbrace  P_{\theta}, \phi \rbrace =\displaystyle \frac{qp}{(p^{2}\sinh^{2}\eta+q^{2})},  \\ \nonumber
	&& \lbrace P_{\theta}, \beta \rbrace =\displaystyle \frac{2\sin\theta}{( p^{2}\sinh^{2}\eta+q^{2})}\frac{\left(p\sinh^{2}\eta P_{\theta}+qP_{\phi}\right)}{(\cosh\eta-\cos\theta)}, 
	 \quad \lbrace \phi, P_{\phi} \rbrace =\displaystyle \frac{p^{2}\sinh^{2} \eta}{(p^{2}\sinh^{2} \eta + q^{2})}, \\ \nonumber
	&&  \lbrace P_{\theta}, \rho \rbrace = \displaystyle -\frac{ma^{2}}{(\cosh\eta-\cos\theta)^2}\frac{p\sinh^{2} \eta}{(p^{2}\sinh^{2}\eta+q^{2})}, \quad   \lbrace \theta, \beta \rbrace = -\displaystyle\frac{p\sinh^{2}\eta}{(p^{2}\sinh^{2}\eta+q^{2})}, \\ \nonumber
	&&	\lbrace  P_{\phi}, \rho \rbrace = - \displaystyle \frac{ma^{2}}{(\cosh\eta - \cos\theta)^{2}}\frac{q \sinh^{2} \eta}{(p^{2}\sinh^{2} \eta + q^{2})}, \quad \lbrace \phi, \beta \rbrace = -\displaystyle \frac{q}{(p^{2}\sinh^{2} \eta + q^{2})},   \\ \nonumber
&&\lbrace  \beta, \rho \rbrace = - \displaystyle \frac{ma^{2}}{(\cosh\eta - \cos\theta)^{2}}\frac{\sinh^{2} \eta}{(p^{2}\sinh^{2} \eta + q^{2})}.
	\end{eqnarray*}
The remaining brackets turn out to be zero. The basic brackets among the phase space variables obtained via Faddeev-Jackiw formalism are found to be coinciding with the corresponding Dirac brackets ({\it{cf.}} Appendix A for explicit calculation of Dirac brackets). Thus, through the medium of basic brackets obtained for the system of particle on a torus knot, we can quantize the theory with the aid of following relation ({\it{cf.}} \cite{int10})
\begin{equation}
\lbrace \zeta_{i}^{(2)}, \zeta_{j}^{(2)} \rbrace = -\frac{i}{\hbar}\left[ \hat{\zeta}_{i}^{(2)}, \hat{\zeta}_{j}^{(2)} \right],
\end{equation}
where $\hat{\zeta}_{i}^{(2)}$ and $\hat{\zeta}_{j}^{(2)}$ represent the operator notation of any general variables in the set of  second-iterated symplectic variables.

\section{Symplectic gauge invariant formalism}
\label{sect:3}
The investigation of a particle on a torus knot via Faddeev-Jackiw formalism affirmed that it is a gauge non-invariant theory.  In this section, we reformulate the same as a gauge invariant theory through the implementation of symplectic gauge invariant formalism \cite{5,6,toric}. 

\subsection{Reformulation as a gauge theory}
\label{sect:3.1}
In order to accomplish the above mentioned goal, the key idea is to enlarge the original phase space by introducing Wess-Zumino variable. To be specific, we introduce a new arbitrary function $G$, which depends upon original phase space variables and Wess-Zumino variable, say $\alpha$, as given below 
\begin{equation}
G(\eta, P_{\eta}, \theta, P_{\theta}, \phi, P_{\phi}, \alpha) \;=\; \displaystyle\sum_{n=0}^{\infty} \mathcal{G}^{(n)}(\eta, P_{\eta}, \theta, P_{\theta}, \phi, P_{\phi}, \alpha),
\end{equation}
where $\mathcal{G}^{(n)}$ is a $n^{th}$ order term in Wess-Zumino variable $\alpha$ and the function $G$ satisfies following boundary condition
\begin{equation}
G(\eta, P_{\eta}, \theta, P_{\theta}, \phi, P_{\phi}, \alpha = 0) \;=\; \mathcal{G}^{(0)} \;=\; 0.
\end{equation}
Thus, working within symplectic gauge invariant formalism and introducing the new function $G$ into  {\eqref{L_1}, the resulting Lagrangian has following form 
\begin{equation} \label{L_f_1}
\tilde{L}_{f}^{(1)} \;=\; \dot{\eta}P_\eta + \dot{\theta}P_\theta + \dot{\phi}P_\phi +\dot{\beta}(p\theta+q\phi) - \tilde{V}^{(1)},
\end{equation}
where the symplectic potential $\tilde{V}^{(1)}$ is given by
\begin{equation} \label{new_sym_pot}
\tilde{V}^{(1)} \;=\; \frac{(\cosh\eta - \cos\theta)^2}{2ma^2}\left[ P_\eta^2 + P_\theta^2 + \frac{P_{\phi}^2}{\sinh^2 \eta} \right] - G.
\end{equation}
The extended set of symplectic variables can be identified as $\tilde{\zeta}^{(1)} = \{\eta, P_\eta, \theta, P_\theta, \phi, P_\phi, \beta, \alpha \}$ and the corresponding canonical one-forms are given as
\begin{eqnarray}
& \tilde{a}^{(1)}_{\eta}\;=\;P_\eta, \quad  \tilde{a}^{(1)}_{\theta}\;=\;P_{\theta}, \quad  \tilde{a}^{(1)}_{\phi}\;=\;P_{\phi},  & \nonumber \\  
& \tilde{a}^{(1)}_{\beta} \;=\; p\theta+q\phi, \quad \tilde{a}^{(1)}_{P_{\eta}}\;=\; \tilde{a}^{(1)}_{P_{\theta}} \;=\; \tilde{a}^{(1)}_{P_{\phi}} \;=\; \tilde{a}^{(1)}_{\alpha} \;=\; 0.&
\end{eqnarray}
With this input the symplectic matrix $\tilde{f}_{ij}^{(1)}$ takes the following form  
\begin{eqnarray} \label{f_singular}
\tilde{f}_{ij}^{(1)} \;=\;
\begin{pmatrix}
0 & -1 &0 &0 &0 &0 &0 &0\\
1 &0 &0 &0 &0 &0 &0 &0\\
0 &0 &0 &-1 &0 &0 &p &0\\
0 &0 &1 &0 &0 &0 &0 &0\\
0 &0 &0 &0 &0 &-1 &q &0\\
0 &0 &0 &0 &1 &0 &0 &0\\
0 &0 &-p &0 &-q &0 &0 &0 \\
0 &0 &0 &0 &0 &0 &0 &0
\end{pmatrix}.
\end{eqnarray}
This symplectic matrix $\tilde{f}_{ij}^{(1)}$ is singular and has a zero-mode of following form
\begin{equation} \label{nu_gauge}
(\tilde{\nu}^{(1)})^{T} \;=\; 
\begin{pmatrix}
0 &0 &0 &p &0 &q &1 &1
\end{pmatrix}.
\end{equation}
To reformulate the system as a gauge theory by symplectic gauge invariant formalism, after incorporating a new arbitrary function $G$ into the first-iterated Lagrangian, we impose the condition that no further constraints are generated by the zero-mode of the corresponding singular symplectic matrix. As a consequence, the contraction of zero-mode $(\tilde{\nu}^{(1)})^{T}$ with the gradient of new symplectic potential ($\tilde{V}^{(1)} = V^{(1)}-G$), must vanish identically as given below
\begin{equation} \label{cond}
(\tilde{\nu}^{(1)})^{T} \frac{\partial\tilde{V}^{(1)}}{\partial \tilde{\zeta}^{(1)}} \;=\; (\tilde{\nu}^{(1)})^{T} \Big( \frac{\partial V^{(1)}}{\partial \tilde{\zeta}^{(1)}} - \displaystyle\sum_{n=0}^{\infty} \frac{\partial \mathcal{G}^{(n)}}{\partial \tilde{\zeta}^{(1)}}\Big) \;=\; 0.
\end{equation}
Thus, from the above expression, we deduce the explicit forms of various order correction terms of arbitrary function $G$. For this, we take the aid of condition that the above expression ({\it{cf.}} \eqref{cond}) must vanish identically null for each power of Wess-Zumino variable. In order to determine the first-order correction in $\alpha$ ({\it{i.e.}} $\mathcal{G}^{(1)}$), considering the terms independent of Wess-Zumino variable and then, by integrating the resultant equation, we compute $\mathcal{G}^{(1)}$ as 
\begin{equation} \label{g_alpha}
\mathcal{G}^{(1)} \;=\; \frac{(\cosh \eta - \cos\theta)^{2}}{ma^{2}} \left(pP_{\theta}+\frac{qP_{\phi}}{\sinh^{2}\eta} \right) \alpha.
\end{equation}
Now this correction term $\mathcal{G}^{(1)}$ is included into the Lagrangian \eqref{L_f_1} in the following fashion
\begin{equation} \label{L_f_1_1}
\tilde{L}_{f}^{\prime (1)} \;=\; \dot{\eta}P_\eta + \dot{\theta}P_\theta + \dot{\phi}P_\phi +\dot{\beta}(p\theta+q\phi) - \tilde{V}^{\prime(1)},
\end{equation}
where the form of symplectic potential $\tilde{V}^{\prime(1)}$ is given by
\begin{eqnarray}
\tilde{V}^{\prime(1)} = \frac{(\cosh\eta - \cos\theta)^2}{2ma^2}\left[ P_\eta^2 + P_\theta^2 + \frac{P_{\phi}^2}{\sinh^2 \eta} 
- 2\left(pP_{\theta}+\frac{qP_{\phi}}{\sinh^{2}\eta} \right) \alpha \right] .
\end{eqnarray}
When the zero-mode $(\tilde{\nu}^{(1)})^{T}$ is being contracted with the gradient of new symplectic potential $\tilde{V}^{\prime(1)}$, it leads to a non-zero term as follows
\begin{equation}
(\tilde{\nu}^{(1)})^{T} \frac{\partial\tilde{V}^{\prime(1)}}{\partial \tilde{\zeta}^{(1)}} \;=\; - \frac{(\cosh \eta - \cos \theta)^{2}}{ma^{2}} \left[p^{2} + \frac{q^{2}}{\sinh^{2} \eta} \right] \alpha,
\end{equation}
which contradicts the condition that zero-mode should not lead to any new constraints. Thus, we have to calculate the second-order correction in $\alpha$ ({\it{i.e.}} $\mathcal{G}^{(2)}$) which can be carried out by looking for the linear order terms in Wess-Zumino variable in \eqref{cond}. 
The expression for $\mathcal{G}^{(2)}$ can be obtained by integrating the aftermath expression and given as
\begin{equation} \label{g2}
\mathcal{G}^{(2)} \;=\; - \frac{(\cosh \eta - \cos \theta)^{2}}{2ma^{2}} \left[p^{2} + \frac{q^{2}}{\sinh^{2} \eta} \right] \alpha^{2}.
\end{equation}
The second-order correction $\mathcal{G}^{(2)}$ is incorporated into the Lagrangian \eqref{L_f_1_1} in the following manner
\begin{equation} \label{L_f_1_2}
\tilde{L}_{f}^{\prime \prime(1)} \;=\; \dot{\eta}P_\eta + \dot{\theta}P_\theta + \dot{\phi}P_\phi +\dot{\beta}(p\theta+q\phi) - \tilde{V}^{\prime\prime(1)},
\end{equation}
where the form of symplectic potential $\tilde{V''}^{(1)}$ is given by
\begin{eqnarray} \label{V_final}
\tilde{V}^{\prime\prime(1)} & = & \frac{(\cosh\eta - \cos\theta)^2}{2ma^2}\Big[ P_\eta^2 + P_\theta^2 + \frac{P_{\phi}^2}{\sinh^2 \eta} -  2 \left(pP_{\theta}+\frac{qP_{\phi}}{\sinh^{2}\eta} \right) \alpha \nonumber \\
& + & \left(p^{2} + \frac{q^{2}}{\sinh^{2} \eta}\right)  \alpha^{2}\Big].
\end{eqnarray}
Now it is straightforward to check that the zero-mode $(\tilde{\nu}^{(1)})^{T}$ contracting on the gradient 
of potential $\tilde{V}^{\prime\prime(1)}$ generates identically null. In other words, the zero-mode $(\tilde{\nu}^{(1)})^{T}$ does not produce any new constraints and, thus, we can stop the iterative process of determining further correction terms with $n\geq 3$ in $G$. Accordingly, this zero-mode turns out to be the generator of infinitesimal gauge symmetry transformations.

Therefore, the zeroth-iterated symplectic potential $(\tilde{V}^{(0)})$, which can also be identified as the canonical Hamiltonian of the system, is given as
\begin{eqnarray} \label{gauge_invariant_V0}
\tilde{V}^{(0)}  &=& \frac{(\cosh\eta - \cos\theta)^2}{2ma^2}\Big( P_\eta^2 + P_\theta^2 + \frac{P_{\phi}^2}{\sinh^2 \eta} -  2 \left(pP_{\theta}+\frac{qP_{\phi}}{\sinh^{2}\eta} \right) \alpha  \nonumber\\
&+&  \left(p^{2} + \frac{q^{2}}{\sinh^{2} \eta}\right)  \alpha^{2}\Big)  - \lambda(p\theta + q\phi)  \;\equiv \; \tilde{H}.
\end{eqnarray}
Correspondingly, the first-order Lagrangian can be expressed as
\begin{equation} \label{L_f_0}
\tilde{L}_{f}^{(0)} \;=\; \dot{\eta}P_\eta + \dot{\theta}P_\theta + \dot{\phi}P_\phi - \tilde{V}^{(0)}.
\end{equation}
As we have already mentioned, the infinitesimal gauge transformations can be generated from the zero-mode ({\it{cf.}} \eqref{nu_gauge}) in the following fashion
\begin{equation}
\delta \tilde{\zeta}^{(1)} = (\tilde{\nu}^{(1)})^{T} \epsilon,
\end{equation} 
where $\epsilon$ is the infinitesimal parameter of gauge transformations and $\tilde{\zeta}^{(1)}$ represents the extended set of first-iterated symplectic variables. The full set of gauge transformations ($\delta$) are listed below
\begin{equation} \label{gauge_trans}
\delta P_{\theta} = p\epsilon, \quad \delta P_{\phi} = q\epsilon, \quad \delta \lambda = \dot{\epsilon}, \quad \delta \alpha = \epsilon, \quad \delta \eta = \delta P_{\eta} = \delta \theta = \delta \phi =0.
\end{equation}
Under the gauge transformations \eqref{gauge_trans}, the zeroth-iterated Lagrangian \eqref{L_f_0} transforms to a total time derivative as follows
\begin{equation}
\delta \tilde{L}_{f}^{(0)} \;=\; \frac{d}{dt} \big[\epsilon(p\theta + q\phi)\big].
\end{equation}
Thus, we have obtained the gauge invariant Lagrangian \eqref{L_f_0} and corresponding Hamiltonian \eqref{gauge_invariant_V0} within the framework of symplectic gauge invariant formalism.

\subsection{Constraint structure and gauge symmetries}
\label{sect:3.2}
In this section, we wish to deduce the constraint structure by employing Dirac's formalism. For this, we identify the canonical conjugate momenta corresponding to $\lambda$ and Wess-Zumino variable $\alpha$ as  primary constraints. These primary constraints $(\psi_1, \bar\psi_1)$ generate two constraint chains. The first chain is as given below
\begin{equation}
\psi_{1} \;=\; P_{\lambda} \approx 0, \quad \psi_{2} \;=\; (p\theta + q\phi) \approx 0,
\end{equation}
and the second constraint chain is listed as
\begin{equation}
\bar{\psi}_{1} = P_{\alpha} \approx 0, \quad 
\bar{\psi}_{2} = \displaystyle\frac{(\cosh \eta - \cos\theta)^{2}}{ma^{2}} \left(pP_{\theta}+\frac{qP_{\phi}}{\sinh^{2}\eta} -  \left(p^{2} + \frac{q^{2}}{\sinh^{2} \eta} \right) \alpha \right) \approx 0 .  
\end{equation}
Here, $P_{\alpha}$ is canonical conjugate momentum to Wess-Zumino variable $\alpha$. At this juncture, we would like to mention that in the first constraint chain the consistency condition of secondary constraint $\psi_{2}$ leads to tertiary constraint $\psi_{3}= \displaystyle\frac{(\cosh \eta - \cos\theta)^{2}}{ma^{2}} \Big(pP_{\theta}+\frac{qP_{\phi}}{\sinh^{2}\eta} -  \big(p^{2} + \frac{q^{2}}{\sinh^{2} \eta} \big) \alpha \Big) \approx 0, $ which is equal to $\bar{\psi_{2}}$ in the second constraint chain. As $\psi_3$ is not unique, so it is possible to put this constraint in one of the chains and simultaneously close the other one \cite{mit1}. Thus, we have retained this constraint in the second chain and at the same time terminated the first chain. 

It is worthwhile to point out that $\psi_1$ has vanishing Poisson brackets with rest of the constraints. Furthermore, there exist non-vanishing Poisson brackets among other constraints such as $\left\lbrace \psi_{2}, \bar{\psi}_{2}  \right\rbrace = \displaystyle\frac{(\cosh \eta - \cos \theta)^{2}}{ma^{2}} \left(p^{2} + \frac{q^{2}}{\sinh^{2} \eta} \right) 
= \left\lbrace \bar{\psi}_{1},\bar{\psi}_{2} \right\rbrace $, which indicate that there is a mixture of both first-class and second-class constraints. So, we split the constraint chains into a set of first-class and second-class constraints through constraint combination \cite{int6,5}. As a result, the set of first-class constraints is
\begin{equation} \label{firstclass_f}
\chi_{1} = P_{\lambda}, \quad
\chi_{2} = p\theta + q\phi - P_{\alpha},
\end{equation}
whereas the set of second-class constraints turns out to be
\begin{equation}
\bar{\psi}_{1} = P_{\alpha}, \quad
\bar{\psi}_{2} = \displaystyle\frac{(\cosh \eta - \cos\theta)^{2}}{ma^{2}} \left(pP_{\theta}+\frac{qP_{\phi}}{\sinh^{2}\eta} 
-  \left(p^{2} + \frac{q^{2}}{\sinh^{2} \eta} \right) \alpha \right).
\end{equation}
Now, the second-class constraints are assumed in a strong way, with this, the Hamiltonian in \eqref{gauge_invariant_V0} is reduced to the following form \cite{gauge_Henn}
\begin{eqnarray} \label{H'_fin}
\tilde{H'} &=& \frac{(\cosh\eta - \cos\theta)^2}{2ma^2}\left( P_\eta^2 + \frac{\left( qP_{\theta}- pP_{\phi}\right)^{2}}{\left( p^{2}\sinh^{2}\eta + q^{2}\right)} \right) - \lambda(p\theta + q\phi).  
\end{eqnarray}
At this juncture, we would like to point out that in the symplectic gauge invariant formalism even though we begin with the inclusion of Wess-Zumino variable by enlarging the phase space, we obtained the form of the Hamiltonian without the presence of any additional variables. Thus, we have reformulated the original gauge non-invariant system as a gauge theory which is free from Wess-Zumino variable.

We now wish to disclose the gauge symmetries of the newly reformulated gauge invariant theory of a particle on a torus knot. For this purpose, we identify the first-order Lagrangian corresponding to Hamiltonian \eqref{H'_fin} as 
\begin{equation} \label{L'_final}
\tilde{L}^\prime_f =  \dot{\eta}P_\eta + \dot{\theta}P_\theta + \dot{\phi}P_\phi  - \frac{(\cosh\eta - \cos\theta)^2}{2ma^2}\left( P_\eta^2 + \frac{\left( qP_{\theta}- pP_{\phi}\right)^{2}}{\left( p^{2}\sinh^{2}\eta + q^{2}\right)} \right)  
+ \lambda(p\theta + q\phi) .
\end{equation}
From the above Lagrangian, we derive the constraints via Dirac formalism as 
\begin{equation} \label{first_generator}
\tilde \chi_{1} = P_{\lambda} \approx 0, \quad
\tilde \chi_{2} = (p\theta + q\phi) \approx 0,
\end{equation}
where the constraints $\tilde \chi_{1}$ and $\tilde \chi_{2}$ are first-class in nature. Consequently, the set of first-class constraints acts as the generator of gauge symmetry and we derive the explicit form of infinitesimal gauge symmetry transformations as
\begin{equation} \label{gauge_trans_firstclass}
\delta P_{\theta} = p\kappa, \quad \delta P_{\phi} = q\kappa, \quad \delta \lambda = \dot{\kappa}, \quad \delta [\eta, P_{\eta}, \theta, \phi]  =0,
\end{equation}
where $\kappa$ represents the infinitesimal gauge parameter. Under the gauge transformations in \eqref{gauge_trans_firstclass}, the first-order Lagrangian in \eqref{L'_final} transforms 
\begin{equation}
\delta \tilde{L}^\prime_f \; = \; \frac{d}{dt} \Big[ \kappa(p\theta + q\phi) \Big],
\end{equation}
into a total time derivative. Thus, by employing symplectic gauge invariant formalism, we have reformulated the system of  particle constrained to move on a torus knot as a gauge theory. 

Before ending this section, a few remarkable features of the newly constructed gauge theory are in order; first,  the expression of first-class constraints preserve the same form as of the original gauge non-invariant theory. Second, the gauge invariant theory is found to be independent of any additional variables even though the symplectic gauge invariant formalism employs a methodology of enlarging the phase space with Wess-Zumino variable. Finally, it is worthwhile to mention that the reformulated gauge theory is {\it physically} equivalent to the original gauge non-invariant one as far as the dynamics and physical degrees of freedom are concerned. To be precise, the equations of motion determined from the newly reformulated gauge theory coincide with Euler-Lagrange equations obtained from the initial gauge non-invariant system (see below, Appendix B for details). Moreover, in the calculation of physical degrees of freedom, taking the case of original gauge non-invariant system, one first-class and two second-class constraints produce one degree of freedom \cite{gauge_Henn}. Likewise, the system of gauge invariant theory with two first-class constraints also provide one degree of freedom. Thus, symplectic gauge invariant formalism produces the same dynamics and preserve the physical degrees of freedom for the reformulated  gauge theory.

\section{(Anti-)BRST symmetry transformations}
\label{sect:4}
Now we investigate the reformulated gauge theory of a particle on a torus knot by means of BRST formalism which benefits in achieving the gauge fixing without breaking underlying symmetry. For this purpose, we introduce anti-commuting (anti-)ghost variables $(\bar c)c$, satisfying $c^{2} = 0 = \bar{c}^{2}, \; c\bar{c}+\bar{c}c=0$ to replace the time dependent gauge parameter. 
Thus, the BRST symmetry transformations $(s_{b})$ can be written as 
\begin{eqnarray} \label{brst}
	 &s_{b}\lambda \;=\; \dot{c}, \quad s_{b} P_{\theta} \;=\; cp, \quad s_{b} P_{\phi} \;=\; cq, \quad	s_{b}\bar{c} \;=\; ib, & \\ \nonumber
& s_{b} c \;=\; 0, \quad s_{b} b \;=\; s_{b} \eta \;=\;  s_{b} P_{\eta} \;=\;  s_{b} \theta \;=\;   s_{b} \phi \;=\;  0,&
\end{eqnarray}
where $b$ is Nakanishi-Lautrup auxiliary variable. The BRST invariant first-order Lagrangian $(L_b)$ can be constructed by adding a BRST invariant function to \eqref{L'_final} as 
\begin{eqnarray} \label{L_brst}
 L_b =  \tilde{L}^\prime_f +  s_{b}\left[ i\bar{c}\left( \dot{\lambda} +  P_\theta + P_\phi  - \frac{1}{2}b \right)\right].
\end{eqnarray}
It can be rewritten in its full blaze of glory as 
\begin{eqnarray} \label{L_brst_1}
 L_b &=& \dot{\eta}P_\eta + \dot{\theta}P_\theta + \dot{\phi}P_\phi  - \frac{(\cosh\eta - \cos\theta)^2}{2ma^2}\Big( P_\eta^2 + \frac{\left( qP_{\theta}- pP_{\phi}\right)^{2}}{\left( p^{2}\sinh^{2}\eta + q^{2}\right)} \Big)  \\ \nonumber 
&+&  \lambda(p\theta + q\phi)  - b (\dot{\lambda} + P_\theta + P_\phi) + \frac{1}{2}b^{2}+ i \dot{\bar{c}}\dot{c} - i (p + q) \bar c c.
\end{eqnarray}
Here the Nakanishi-Lautrup auxiliary variable is used to linearize the gauge fixing term $\displaystyle-\frac{1}{2}\left[\dot{\lambda} + P_\theta + P_\phi \right]^{2}$. Further, the BRST invariant first-order Lagrangian \eqref{L_brst_1} is also found to be invariant under another set of symmetries, known as anti-BRST symmetries ($s_{ab}$), of the following form  
\begin{eqnarray} \label{antibrst}
	&s_{ab}\lambda \;=\; \dot{\bar c}, \quad s_{ab} P_{\theta} \;=\; {\bar c}p, \quad s_{ab} P_{\phi} \;=\; {\bar c}q, \quad	s_{ab}c \;=\; -ib, & \\ \nonumber
&s_{ab} \bar{c} \;=\; 0, \quad s_{ab} b \;=\; s_{ab} \eta \;=\;  s_{ab} P_{\eta} \;=\;  s_{ab} \theta \;=\;  s_{ab} \phi \;=\;  0.&
\end{eqnarray}
These (anti-)BRST symmetry transformations are off-shell nilpotent ({\it{i.e.}} $s_{b}^{2} = 0 = s_{ab}^{2} $) and absolutely anti-commuting ({\it{i.e.}} $\left\lbrace s_{b}, s_{ab} \right\rbrace  =  s_{b}s_{ab} + s_{ab}s_{b} = 0$) in nature. The Hamiltonian corresponding to $L_b$ can be expressed as 
\begin{eqnarray}
H_{b} = \dot{\eta}P_{\eta} + \dot{\theta}P_{\theta} + \dot{\phi}P_{\phi} + \dot{\lambda}P_{\lambda} + \dot{c}\Pi_{c} + \dot{\bar{c}}\Pi_{\bar{c}} - L_{b}, 
\end{eqnarray}
with $P_{\lambda} = -b$, $\Pi_{c} = -i\dot{\bar{c}}$ and $\Pi_{\bar{c}} = i\dot{c}$, we obtain
\begin{eqnarray}
 H_{b} &=&  \frac{(\cosh\eta - \cos\theta)^2}{2ma^2}\Big( P_\eta^2 + \frac{\left( qP_{\theta}- pP_{\phi}\right)^{2}}{\left( p^{2}\sinh^{2}\eta + q^{2}\right)} \Big)  -i \Pi_{\bar{c}}\Pi_{c} \nonumber \\
 &-& \lambda(p\theta + q\phi) - \frac{1}{2}P_{\lambda}^{2} - P_\lambda (P_\theta + P_\phi) + i (p + q) {\bar c} c. 
 \end{eqnarray} 
The conserved (anti-)BRST charges $Q_{(a)b}$ can be derived as
\begin{equation}
	Q_{b} = \dot{c}P_{\lambda} - c(p\theta+q\phi), \quad Q_{ab}= \dot{\bar{c}}P_{\lambda} -\bar{c}(p\theta+q\phi).
\end{equation}
It is worth mentioning that the (anti-)BRST charges are nilpotent of order two ({\it{i.e.}}  $Q_{b}^{2} = 0 = Q_{ab}^{2}$) and absolutely anti-commuting ($\lbrace Q_{b}, Q_{ab}\rbrace = Q_{b}Q_{ab}+Q_{ab}Q_{b}=0$) in nature. In addition, the conservation of charges $Q_{(a)b}$ can be verified with the aid of Euler-Lagrange equations of motion corresponding to the first-order Lagrangian ({\it{cf.}} \eqref{L_brst_1}) as given below 
\begin{eqnarray}
&& \dot{\theta} = b + \displaystyle \frac{(\cosh \eta - \cos \theta)^{2}}{ma^{2}} \frac{q\left(q P_{\theta} - pP_{\phi}\right) }{\left( p^{2}\sinh^{2}\eta + q^{2}\right) }, \qquad \dot{\eta} = \displaystyle\frac{(\cosh \eta - \cos \theta)^{2}}{ma^{2}}P_{\eta},
  \\ \nonumber
&& \dot{\phi} = b -\displaystyle \frac{(\cosh \eta - \cos \theta)^{2}}{ma^{2}} \frac{p\left(q P_{\theta} - pP_{\phi}\right) }{\left( p^{2}\sinh^{2}\eta + q^{2}\right) }, \qquad \dot{b} = - (p\theta+q\phi), \\ \nonumber
&& b = \dot \lambda + P_\theta + P_\phi, \qquad \ddot{c} + (p + q) c = 0, \qquad \ddot{\bar{c}} + (p + q) \bar c = 0, \\  \nonumber
&&  \dot{P_{\phi}} = \lambda q, \qquad  \dot{P_{\theta}} = \displaystyle \lambda p - \frac{(\cosh\eta - \cos\theta)\sin\theta}{ma^2}\Big[ P_\eta^2 + \frac{\left( qP_{\theta}- pP_{\phi}\right)^{2}}{\left( p^{2}\sinh^{2}\eta + q^{2}\right)}  \Big],  \\ \nonumber 
&&  \dot{P}_{\eta} = - \frac{(\cosh\eta - \cos\theta)}{ma^2}\sinh\eta\Big[ P_\eta^2 
+ \frac{\left( qP_{\theta}- pP_{\phi}\right)^{2}}{\left( p^{2}\sinh^{2}\eta + q^{2}\right) } \\ \nonumber 
&& -  \frac{ (\cosh\eta - \cos\theta) \cosh\eta  \left(qP_{\theta}-pP_{\phi}\right)^{2} p^{2}}{\left(p^{2}\sinh^{2}\eta+q^{2}\right)^2}  \Big]. 
\end{eqnarray}
Further, we can verify that the conserved nilpotent charges $Q_{(a)b}$ are the generators of (anti-)BRST transformations ({\it{cf.}} \eqref{brst}, \eqref{antibrst}) from the following relation 
\begin{equation}
	s_{(a)b}\psi = i\left[ Q_{(a)b}, \psi \right] _{\pm},
\end{equation}
 by making use of (anti-)commutation relations: $ \left[\eta, P_{\eta}\right] = \left[\theta, P_{\theta}\right] = \left[\phi, P_{\phi}\right] = \left[ \lambda, P_{\lambda}\right] = i, \; \left\lbrace \dot{\bar{c}},c \right\rbrace =1,$ and $ \left\lbrace \dot{c}, \bar{c} \right\rbrace =-1 $. In the above, $\psi$ is any generic variable of the system and $\pm$ represent the  (anti-)commutation relation. 
 The physicality criteria implies that these (anti-)BRST charges annihilate the physical subspace 
  $\left| \varPsi \right\rangle_{phys}$, which leads to 
 \begin{equation}
	Q_{(a)b} \left| \varPsi \right\rangle_{phys}   =0 \quad  \Longrightarrow \quad P_{\lambda}\left| \varPsi\right\rangle_{phys}  = 0, \quad (p\theta+q\phi)\left| \varPsi \right\rangle _{phys} =0.
\end{equation}
This shows the consistency with the Dirac quantization scheme as the physical states in the theory are annihilated by the first-class constraints.

\section{Conclusions}
\label{sect:5}
In the present endeavor, we have quantized the system of particle constrained to move on a torus knot by the means of Faddeev-Jackiw formalism. We have also derived the constraint structure and further found that the two-form symplectic matrix becomes non-singular after incorporating all the constraints into the canonical sector of the first-order Lagrangian. This shows that the underlying theory is gauge non-invariant and from the inverse of the non-singular symplectic matrix, we have obtained all the basic brackets.

We have further employed symplectic gauge invariant formalism to reformulate this gauge non-invariant theory into a gauge invariant one. To implement this formalism, we have enlarged the original phase space by introducing a new variable - the Wess-Zumino variable. In the procedure, we have included an arbitrary function depending upon the original phase space variables and Wess-Zumino variable into the first-iterated Lagrangian. Subsequently, by imposing the condition that the zero-mode of corresponding singular symplectic matrix does not lead to further constraints, we have found the explicit form of the arbitrary function. The analysis of constraint structure for the newly constructed theory have revealed the presence of a mixture of both the first-class and second-class constraints. Later, through constraint combination, the first-class constraints have been separated out and second-class constraints are assumed in a strong way. This, in turn, produced a Hamiltonian (and corresponding Lagrangian)  which is free from any Wess-Zumino variables. Furthermore, we have demonstrated that the thus obtained, first-order Lagrangian remains invariant under the transformations generated by the first-class constraints.

Finally, we have deduced BRST invariant first-order Lagrangian and established the off-shell nilpotent and absolutely anti-commuting (anti-)BRST symmetries corresponding to the newly reformulated gauge theory. We have also derived the (anti-)BRST charges which are shown to be nilpotent of order two and absolutely anti-commuting in nature. Further, these conserved (anti-)BRST charges have been depicted as the generators of the corresponding (anti-)BRST symmetries. Additionally, the physicality condition of dynamically stable subspace shown that the first-class constraints annihilate the physical states of the system and is consistent with the Dirac quantization procedure.
For future endeavor, we wish to explore the possibility of a particle on torus knot as a model for Hodge theory and aim to establish the physical realizations of cohomological operators within the BRST formalism \cite{torus_knot}. \\

\vspace{4mm}
\noindent 
{\large \bf Acknowledgments:} The support from FRG scheme of National Institute of Technology Calicut is thankfully acknowledged.

\appendix

\section{Hamiltonian analysis: Dirac brackets}
As  we have already seen in Section 2, the Dirac Hamiltonian analysis for the system of particle on a torus knot revealed the presence of one first-class ($\varphi_{1}$) and two second-class constraints ($\varphi_{2}, \varphi_{3}$). Therefore, from the second-class constraints  $\varphi_{2} = (p\theta+q\phi) \approx 0$ and $\varphi_{3}  =  \frac{(\cosh\eta - \cos\theta)^{2}}{ma^{2}} \left( pP_\theta + \frac{qP_{\phi}}{\sinh^{2}\eta} \right) \approx 0$, we can construct a $2 \times 2$ anti-symmetric matrix $C_{ij} = \left\lbrace \varphi_{i}, \varphi_{j} \right\rbrace $, with $i, j = 2, 3$, as
\begin{eqnarray}
	C_{ij} \;=\;
	\begin{pmatrix}
      0 &  \frac{(\cosh\eta - \cos\theta)^{2}}{ma^{2}} \left( p^{2} + \frac{q^{2}}{\sinh^{2}\eta} \right) \\
		-\frac{(\cosh\eta - \cos\theta)^{2}}{ma^{2}} \left( p^{2} + \frac{q^{2}}{\sinh^{2}\eta} \right) &0 
	\end{pmatrix}.\;
\end{eqnarray}
This is a non-singular matrix and from the inverse of $C_{ij}$ we can compute Dirac brackets. For any dynamical variables, say  $A$ and $B$, of a theory the Dirac brackets $(\left[ A, B \right]_{D})$ can be given as 
\begin{equation}
\left[ A, B \right]_{D} = \left[ A, B \right] - \left[ A, \varphi_{i}\right]C_{ij}^{-1} \left[\varphi_{j}, B \right],
\end{equation}
where $\left[ A, B \right]$ defines the Poisson bracket between the dynamical variables and $C_{ij}^{-1}$ represents the inverse of the anti-symmetric matrix $C_{ij}$, which, in this case, turns out to be 
\begin{eqnarray}
	C_{ij}^{-1} =
	\begin{pmatrix}
		0 &  - \frac{ma^{2}}{(\cosh\eta - \cos\theta)^{2}} \left( p^{2} + \frac{q^{2}}{\sinh^{2}\eta} \right)^{-1} \\
	  \frac{ma^{2}}{(\cosh\eta - \cos\theta)^{2}} \left( p^{2} + \frac{q^{2}}{\sinh^{2}\eta} \right)^{-1} &0
	\end{pmatrix}.\;\;\;\;\;\;
\end{eqnarray}
Thus, Dirac brackets can be worked out and the non-zero ones are listed below
\begin{eqnarray}
	&& \left[\eta, P_{\eta} \right]_{D} = 1, \quad \left[ \theta, P_{\phi}  \right]_{D} = \displaystyle -\frac{qp\sinh^{2}\eta}{(p^{2}\sinh^{2}\eta+q^{2})}, \quad \left[ \phi, P_{\phi}  \right]_{D} =\displaystyle \frac{p^{2}\sinh^{2} \eta}{(p^{2}\sinh^{2} \eta + q^{2})},  \\ \nonumber
	&&  \left[ \theta, P_{\theta} \right]_{D} = \displaystyle \frac{q^{2}}{(p^{2}\sinh^{2}\eta+q^{2})}, \qquad \left[  P_{\theta}, \phi  \right]_{D} =\displaystyle \frac{qp}{(p^{2}\sinh^{2}\eta+q^{2})}, \\ \nonumber 
	&&  \left[ P_{\eta}, P_{\theta} \right]_{D} = \displaystyle \frac{2p}{(p^{2}\sinh^{2}\eta+q^{2})}\left[ \frac{\sinh\eta}{(\cosh\eta-\cos\theta)} \left(p\sinh^{2}\eta P_{\theta}+qP_{\phi}\right) -q\coth\eta P_{\phi} \right], \\ \nonumber 
	&& \left[ P_{\eta}, P_{\phi} \right]_{D} = \displaystyle \frac{2q}{(p^{2}\sinh^{2}\eta+q^{2})}\left[ \frac{\sinh\eta}{(\cosh\eta-\cos\theta)} \left(p\sinh^{2}\eta P_{\theta}+qP_{\phi}\right) -q\coth\eta P_{\phi} \right], \\ \nonumber
	&& \left[ P_{\theta}, P_{\phi} \right]_{D} = \displaystyle \frac{2q \sin\theta}{(p^{2}\sinh^{2}\eta+q^{2})}\frac{\left(p\sinh^{2}\eta P_{\theta}+qP_{\phi}\right)}{(\cosh\eta-\cos\theta)}. 
\end{eqnarray}
We infer that the above procured Dirac brackets, among the dynamical variables of the theory, coincide with the basic brackets calculated by employing Faddeev-Jackiw formalism ({\it{cf.}} \eqref{FJ_basic brac}). 

\section{On Lagrange multipliers}
In Faddeev-Jackiw formalism, Lagrange multipliers appear as auxiliary coordinates with the purpose of incorporating constraints into the canonical sector of the first-order Lagrangian. Whereas in Dirac formalism these Lagrange multipliers remain absent. Thus, we furnish a {\it{new}} interpretation for these Lagrange multipliers in terms of dynamical coordinates in the system \cite{mFJ3}. For this purpose we make use of symplectic equations of motion of second-iterated Lagrangian  \eqref{L_2}
\begin{eqnarray}
\dot{\zeta}^{(2)j} =(f_{ij}^{(2)})^{-1}\frac{\partial V^{(2)}(\zeta)}{\partial \zeta^{(2)i}}.
\end{eqnarray}
The explicit form of the above relationship is given as
\begin{eqnarray} \label{f_ij_2}
&\begin{pmatrix}
\dot{\eta}\\
\dot{P_{\eta}}\\
\dot{\theta}\\
\dot{P_{\theta}}\\
\dot{\phi}\\
\dot{P_{\phi}}\\
\dot{\beta}\\
\dot{\rho}
\end{pmatrix}=
\begin{pmatrix}
0  & 1 & 0 & 0 & 0 & 0 & 0 & 0\\
-1 & 0 & 0 &\frac{pf_{\eta\rho}}{c} & 0 &\frac{qf_{\eta\rho}}{c} & \frac{f_{\eta\rho}}{c} & 0\\
0   &0 & 0 &\frac{qf_{P_{\phi}\rho}}{c} &0 &-\frac{qf_{P_{\theta}\rho}}{c} &-\frac{f_{P_{\theta}\rho}}{c} & 0\\
0  &-\frac{pf_{\eta\rho}}{c} &-\frac{qf_{P_{\phi}\rho}}{c} &0 &\frac{pf_{P_{\phi}\rho}}{c} &\frac{qf_{\theta\rho}}{c} &\frac{f_{\theta\rho}}{c} &-\frac{p}{c}\\
0 & 0 & 0 &-\frac{pf_{P_{\phi}\rho}}{c} &0 &\frac{pf_{P_{\theta}\rho}}{c} &-\frac{f_{P_{\phi}\rho}}{c} &0\\
0 &-\frac{qf_{\eta\rho}}{c} &\frac{qf_{P_{\theta}\rho}}{c} &-\frac{qf_{\theta\rho}}{c} &-\frac{pf_{P_{\theta}\rho}}{c} &0 &0 &-\frac{q}{c}\\
0 &-\frac{f_{\eta\rho}}{c} &\frac{f_{P_{\theta}\rho}}{c} &-\frac{f_{\theta\rho}}{c} &\frac{f_{P_{\phi}\rho}}{c} &0 &0 &-\frac{1}{c}\\
0 &0 &0 &\frac{p}{c} &0 &\frac{q}{c} &\frac{1}{c} &0
\end{pmatrix}& \\ \nonumber
&\begin{pmatrix}
\displaystyle \frac{(\cosh\eta - \cos\theta)\sinh\eta}{ma^{2}} \left(P_{\eta}^{2}+P_{\theta}^{2} + \frac{P_{\phi}^{2}}{\sinh^{2}\eta} \right) - \frac{(\cosh\eta - \cos\theta)^{2}\cosh\eta P_{\phi}^{2}}{ma^{2}\sinh^{3}\eta}\\
\displaystyle\frac{(\cosh\eta - \cos\theta)^{2}}{ma^{2}}P_{\eta}\\
\displaystyle\frac{(\cosh\eta - \cos\theta)\sin\theta}{ma^{2}} \left(P_{\eta}^{2}+P_{\theta}^{2} + \frac{P_{\phi}^{2}}{\sinh^{2}\eta} \right)\\
\displaystyle\frac{(\cosh\eta - \cos\theta)^{2}}{ma^{2}}P_{\theta} \\
0\\
\displaystyle \frac{(\cosh\eta - \cos\theta)^{2}}{ma^{2}}\frac{P_{\phi}}{\sinh^{2}\eta}\\
0 \\
0
\end{pmatrix}.&
\end{eqnarray}
From the above relationship \eqref{f_ij_2}, it is straightforward to identify the expression for Lagrange multipliers as
\begin{eqnarray}
\dot\rho &=& \displaystyle \frac{\sinh^{2}\eta}{\left(p^{2}\sinh^{2}\eta+q^{2}\right)} \left(pP_{\theta} + \frac{qP_{\phi}}{\sinh^{2}\eta} \right), \\ \nonumber
\dot\beta &=&  \displaystyle \frac{\sinh^{2}\eta}{ma^{2}}\frac{\left(\cosh\eta - \cos\theta\right)}{\left(p^{2}\sinh^{2}\eta+q^{2}\right)}\Big[ p\sin\theta \left(P_{\eta}^{2}+P_{\theta}^{2} + \frac{P_{\phi}^{2}}{\sinh^{2}\eta}\right) \\ \nonumber 
&+& \frac{2q\cosh\eta}{\sinh^{3}\eta}\left(\cosh\eta - \cos\theta\right) P_{\eta}P_{\phi} - 2\left(pP_{\theta} + \frac{qP_{\phi}}{\sinh^{2}\eta} \right) \left[ \sinh\eta P_{\eta} + \sin\theta P_{\theta} \right] \Big] .
\end{eqnarray}
Thus, from the above expressions it is evident that the Lagrange multipliers, which appear as auxiliary variables  in Faddeev-Jackiw formalism, can be expressed in terms of generalized coordinates and momenta. These expressions provide a {\it{new}} interpretation for the Lagrange multipliers.
In our investigation, we have procured all constraints via Faddeev-Jackiw formalism and subsequently quantized the system. Finally, we wish to write down the equations of motion and from the second-iterated symplectic matrix these equations of motion take the following form
\begin{equation}
f_{ij}^{(2)}\dot{\zeta}^{(2)j} = \frac{\partial V^{(2)}(\zeta)}{\partial \zeta^{(2)i}},
\end{equation}
where the explicit expressions are given as
\begin{eqnarray} \label{f2_eom}
&& \displaystyle \dot{P_{\phi}}- q\dot{\beta} \;=\; 0,  
\qquad  p\dot{\theta}+ q\dot{\phi}\;=\; 0,  \qquad  \dot{\theta} + \frac{ (\cosh \eta - \cos \theta)^{2}} {ma^{2}}  \Big(p\dot{\rho} - P_{\theta} \Big) \;=\; 0 ,\\ \nonumber
&& \displaystyle  \dot{\phi} + \frac{ (\cosh \eta - \cos \theta)^{2}} {ma^{2}\sinh^{2} \eta}  \Big(q\dot{\rho} -  P_{\phi} \Big) \;=\; 0,  \qquad \dot{\eta} - \frac{ (\cosh \eta - \cos \theta)^{2}}{ma^{2}} P_{\eta} \;=\; 0,  \\ \nonumber 
&& \displaystyle \dot{P}_{\theta} - p\dot{\beta} - f_{\theta\rho} \dot{\rho} \;=\;- \frac{(\cosh\eta - \cos\theta)\sin\theta} {ma^{2}} \left(P_{\eta}^{2}+P_{\theta}^{2} + \frac{P_{\phi}^{2}}{\sinh^{2}\eta} \right), \\ \nonumber 
&& \displaystyle  \dot{P}_{\eta} -   f_{\eta\rho}\dot{\rho} \;=\; \frac{(\cosh\eta - \cos\theta)^{2}\cosh\eta }{ma^{2}\sinh^{3}\eta} P_{\phi}^{2}  
- \frac{(\cosh\eta - \cos\theta)\sinh\eta}{ma^{2}} \left(P_{\eta}^{2}+P_{\theta}^{2} + \frac{P_{\phi}^{2}}{\sinh^{2}\eta} \right), \\ \nonumber 
&& \dot \eta f_{\eta\rho} + \dot \theta f_{\theta\rho} + \dot P_\theta f_{P_\theta \rho} + \dot P_\phi f_{P_\phi \rho} \;=\; 0.
\end{eqnarray}
The precise forms of $f_{\theta\rho}$, $ f_{\eta\rho}$,  $f_{P_\theta \rho}$ and $ f_{P_\phi \rho}$ are given in \eqref{f_{ij}_forms}. Thus, we have obtained the Euler-Lagrange equations corresponding to the second-iterated Lagrangian.

\end{document}